\documentclass{aa}  

\usepackage{graphicx}
\usepackage{txfonts}

\begin{document}

\title{Radio and optical realizations of celestial reference frames}

\author{S. B. Lambert\inst{1} \and S. Bouquillon\inst{2} \and C. Le Poncin-Lafitte\inst{3} \and C. Barache\inst{2} \and J. Souchay\inst{2}}

\offprints{S. B. Lambert \email{s.lambert@oma.be}}

\institute{Royal Observatory of Belgium, B-1180 Brussels
           \and
           Observatoire de Paris, SYRTE/CNRS UMR8630, F-75014 Paris
           \and
           Lohrmann Observatory, Dresden Technical University, D-01062 Dresden
           }

\date{}

\abstract
{The International Celestial Reference Frame (ICRF, Ma et al. 1998) is currently the best realization of a quasi-inertial reference system. It is based on more than 10 years of cumulated geodetic and astrometric VLBI observations of compact extragalactic objects at centimetric wavelengths.}
{In the perspective of the realization of an accurate optical counterpart of the ICRF using future space astrometry missions like GAIA or SIM, this paper investigates the consistency of celestial reference frames realized through the same subset of compact extragalactic radio sources at optical wavelengths.}
{Celestial reference frames realized in radio wavelengths with the VLBA Calibrator Survey (VCS) data and in optical wavelengths with the Sloan Digital Sky Survey (SDSS) data (DR3 quasar catalogue and DR5) are compared in terms of radio-optical distances between the common sources, global rotation of the axes and offset of the equator.}
{186 sources are cross-identified between the VCS data and the SDSS DR3 quasar catalogue. 796 sources are cross-identified between the VCS data and the SDSS DR5. The accuracy of the SDSS is found around 100~mas, consistently with the value estimated by the astrometric calibration of Pier et al.~(2003). Celestial reference frames realized with the radio and optical coordinates of the cross-identified sources are consistent at the level of 20~mas. A~$-20\pm5$~mas global rotation of the VCS with respect to the SDSS (both releases) sources shows up around the $Y$-axis. This significant effect appears to be on the angle $A_2$ mainly because the SDSS covers a limited area centered in the direction of $-X$ ($\alpha=12$~hours). It is expected to disappear on $A_2$ and show up on $dz$ if the coverage in right ascension were uniform. This effect, statistically significant compared to the VCS error ellipse, is likely due to a non optimal calibration of the SDSS and should be addressed for future tying of radio and optical astrometric data~sets.}
{}


\authorrunning{S. B. Lambert et al.}
\titlerunning{Radio and optical celestial reference frames}

\maketitle

\section{Introduction}

A wide field of research in astronomy, celestial mechanics or geophysics needs the most accurate realization of an inertial celestial reference system. Celestial reference frames were built first through optical telescopes with an accuracy around 0.1~arc second. The development of very long baseline interferometry (VLBI), with observations at centimetric wavelengths, improved considerably the celestial frames, offering the milliarc second accuracy in the late 1980s, and the ability to propose a conventional International Celestial Reference Frame (ICRF, Ma et al.~1998). The ICRF was adopted as the fundamental celestial reference frame at the International Astronomical Union (IAU) 23rd General Assembly, at Kyoto, Japan, in 1997. The ICRF is defined by the radio positions of 212 'defining' compact extragalactic objects, obtained from more that 10 years of VLBI observations (1.6~million pairs of group delays and phase delay rates). It offers a positional accuracy better than the milliarc second, is non-rotating with respect to an inertial frame, and shows little or no time dependency and a good time stability with axes determined at the level of a few tens of micro arc seconds (Ma et al. 1998, Gontier et al. 2001).

The realization of the radio reference frame is still undergoing developments. Recent works by Feissel-Vernier (2003) and Feissel-Vernier et al. (2005, 2006) investigated more stable radio reference frames using a selection of radio sources based on the analysis of time series of radiocenter positions, since radiocenter positions are modified at the milliarc second level over a few years by internal processes in the sources (e.g., jets, see for instance Fey et al. 1997, Feissel et al. 2000). Charlot et al. (2006) investigated the astrometric suitability of radio sources through their intrinsic structure. The IAU Working Group on the Next ICRF, in which two of us (SBL and JS) are involved, has been settled recently at the last IAU General Assembly (Prague, August 2006) and will be chaired by C. Ma (National Aeronautic and Space Administration/Goddard Space Flight Center, Greenbelt, Maryland) with the aim to define a more accurate and more stable conventional celestial radio reference frame to replace the current ICRF.

With the development of optical surveys, quasi-inertial, optical, celestial reference frames will be soon defined with a similar accuracy than the ICRF. The maintenance and the development of the radio reference frame will obviously remain crucial since VLBI is the only technique providing regularly and precisely the orientation of the Earth's crust with respect to space (precession and nutation) and its sidereal rotation (UT1), impacting therefore geosciences, particularly the determination of the Earth's structural parameters (see, e.g., Mathews et al. 2002, Dehant et al. 2003, Feissel-Vernier et al. 2004) or the study of the Earth's rotating fluid core (e.g., Herring et al. 2002, Vondr\'ak et al. 2005, Lambert 2006).

Nevertheless, since an increasing number of optical observations of quasars are provided by ground-based and space surveys, a future work will be to tie the radio reference frame to its optical counterpart in order to align the positions of the optical sources to the ICRF. Such a procedure would permit to assess the quality of the optical catalogues and to remove any systematic effect in them. Another interest is the definition and the maintenance of an optical celestial reference frame and the selection of a large set of astrometric standards among the optical sources in view of future large surveys, especially for the upcoming space astrometry missions like the NASA's SIM (2009, Danner et al. 1999) or ESA's GAIA (2011, Perryman et al. 2001) for which the astrometric accuracy is expected to be close to the VLBI~one (Mignard 2003).

In this study, the coordinates of a set of extragalactic objects (e.g., quasars, BL Lac, AGN) taken from the fifth release of the Very Long Baseline Array (VLBA) Calibrator Survey (Kovalev et al. 2006), are compared to their optical counterparts taken from the Sloan Digital Sky Survey (SDSS) data release~3 quasar catalogue (Schneider et al. 2003) and from the SDSS data release~5 (Adelman-McCarthy et al. 2006). The ability to realize a celestial reference frame at both radio and optical wavelengths is investigated in terms of global rotations and stability of the axes and the pole.

\section{Cross-identification of radio and optical catalogues}

The fifth release of the VLBA Calibrator Survey (denoted as VCS in the following, detailed in Kovalev et al. 2006) data contains 3,357 extragalactic radio sources, observed by the North-American very long baseline array network at 8.3~GHz (X-band) and 2.3~GHz (S-band). The positions were derived from the analysis of ionosphere-free combination of the 22 VCS campaigns 1--5 sessions together with 3,976 dual-frequency multi-baseline 24-hr geodetic VLBI experiments, and using the Calc/Solve geodetic VLBI analysis software package maintained and developped at the NASA/GSFC. To align the celestial frame to the current ICRF, a no-net rotation constraint was applied to the 212 ICRF defining sources (Ma et al. 1998). The mean semimajor axis of the error ellipse of the source coordinates is shown to be close to 5~milliarc seconds (mas) following Petrov et al.~(2006).

\begin{figure}[htbp]
\centering
\includegraphics[width=8cm]{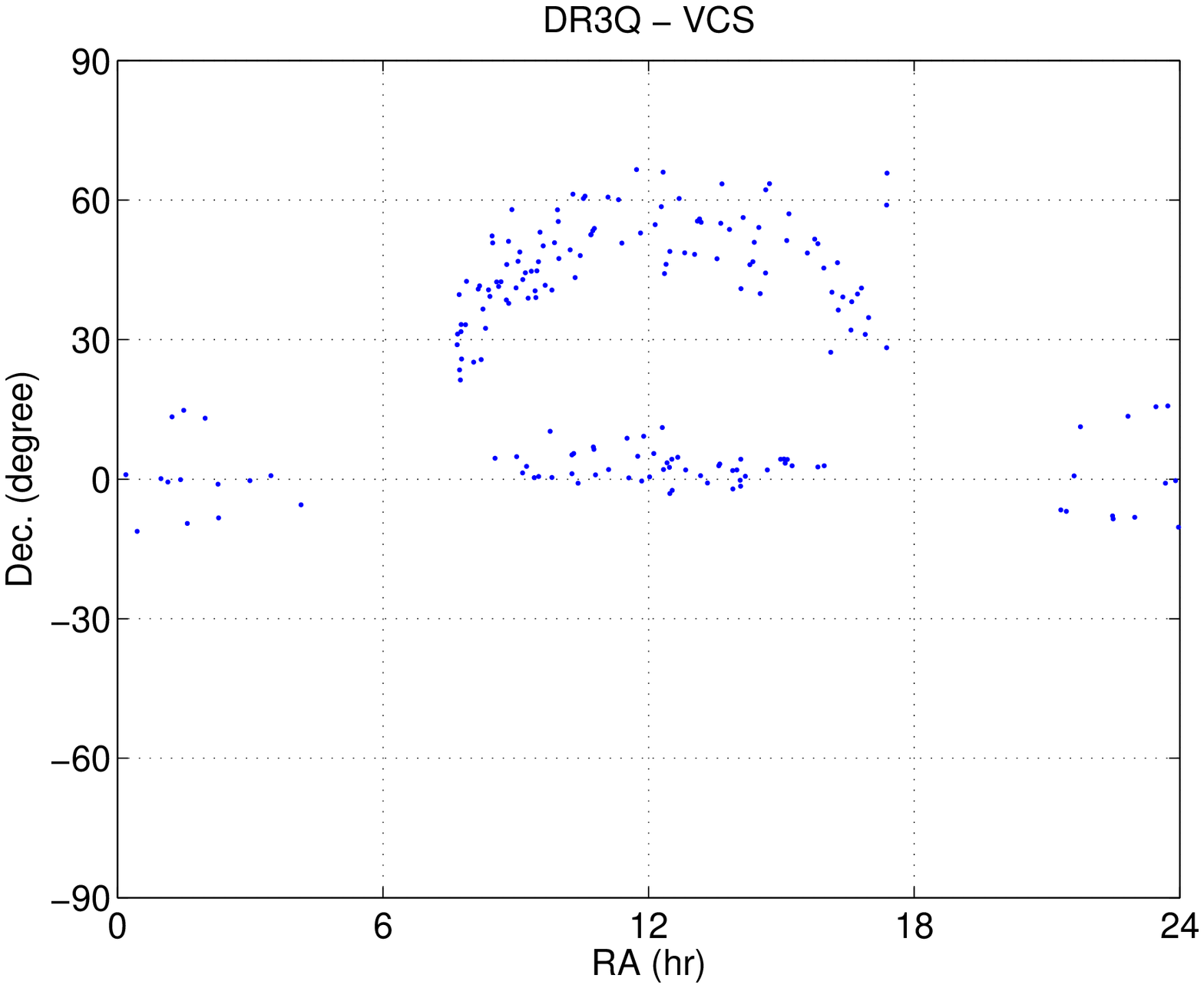}
\includegraphics[width=8cm]{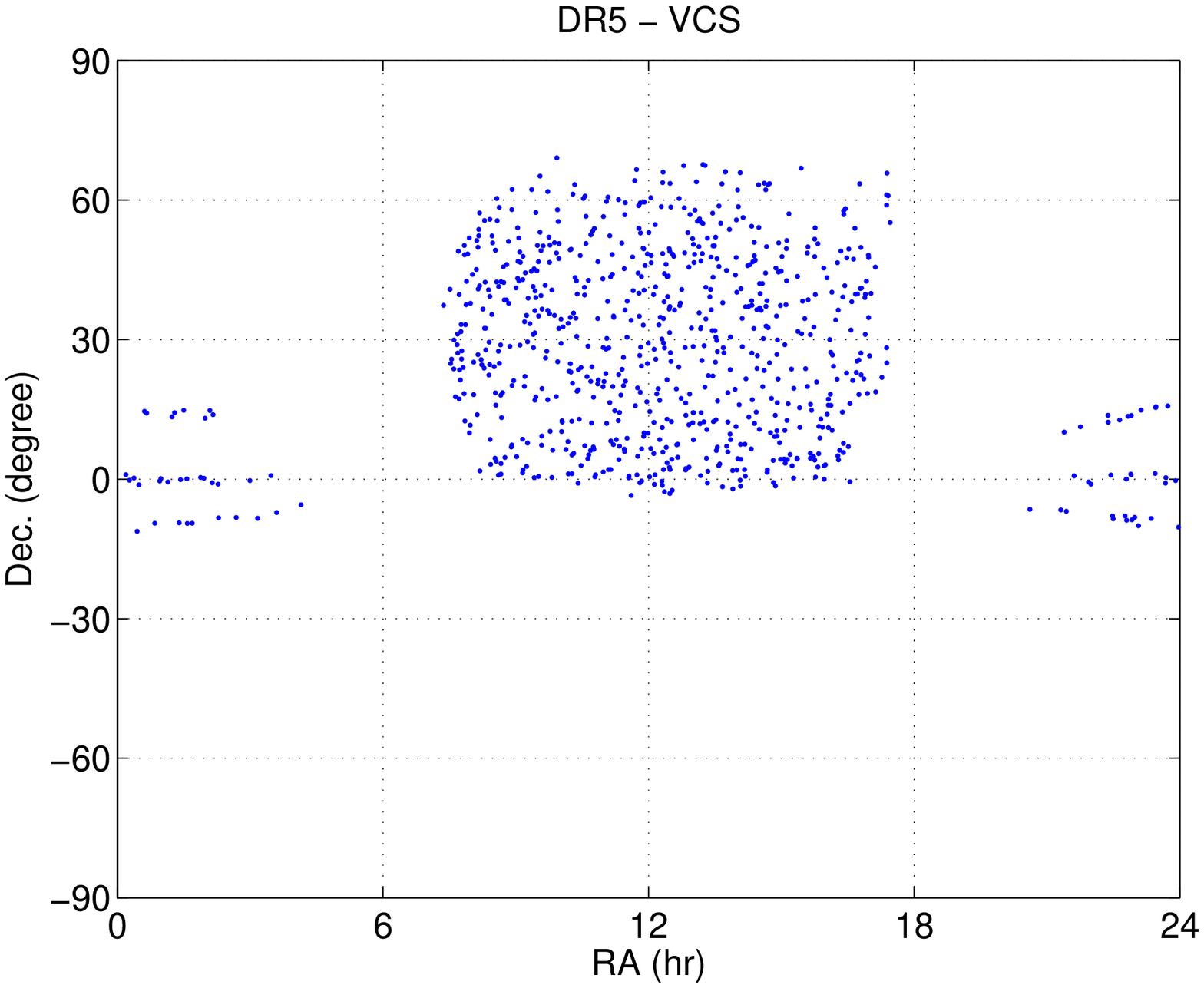}
\caption{Sky distribution of the cross-identified sources between (top) the VCS and the DR3Q, and (bottom) between the VCS and the DR5, both with a radius of 1~arc second.}
\label{fig01}
\end{figure}

The Sloan Digital Sky Survey is covering about one quarter of the sky, observed from a dedicated 2.5-m telescope located at Apache Point, New Mexico. Images are obtained in five broad optical bands (designated $u$, $g$, $r$, $i$, $z$) covering the wavelength range of the CCD response from atmospheric ultraviolet cutoff to the near infrared (see Fukugita et al.~1996 for details). The astrometric calibration (Pier et al. 2003) yields an accuracy per coordinate of 45~mas when reduced against the USNO CCD Astrograph catalogue (UCAC) and 75~mas when reduced against Tycho-2.

We use two data sets from the SDSS. First, we consider the SDSS data release~3 quasar catalogue (Schneider et al.~2003), referred to as DR3Q in the following, consisting in 46,420 objects.  The quasars selection algorithm retains objects intrinsically brighter than $M_i=-22$, assuming a cosmology consistent with the Wilkinson Microwave Anisotropy Probe (WMAP) results, showing at least one emission line with FWHM larger than 1000~km/s, fainter than $m_i=15$ and having a reliable determination of the redshift.

Second, we consider the SDSS data release~5 (DR5, Adelman-McCarthy et al. 2006) yielding some 215 millions objects (most of them are stars) and available through SQL query on the DR5 web site (SDSS DR5~2006).

\begin{figure*}[htbp]
\centering
\includegraphics[width=8cm]{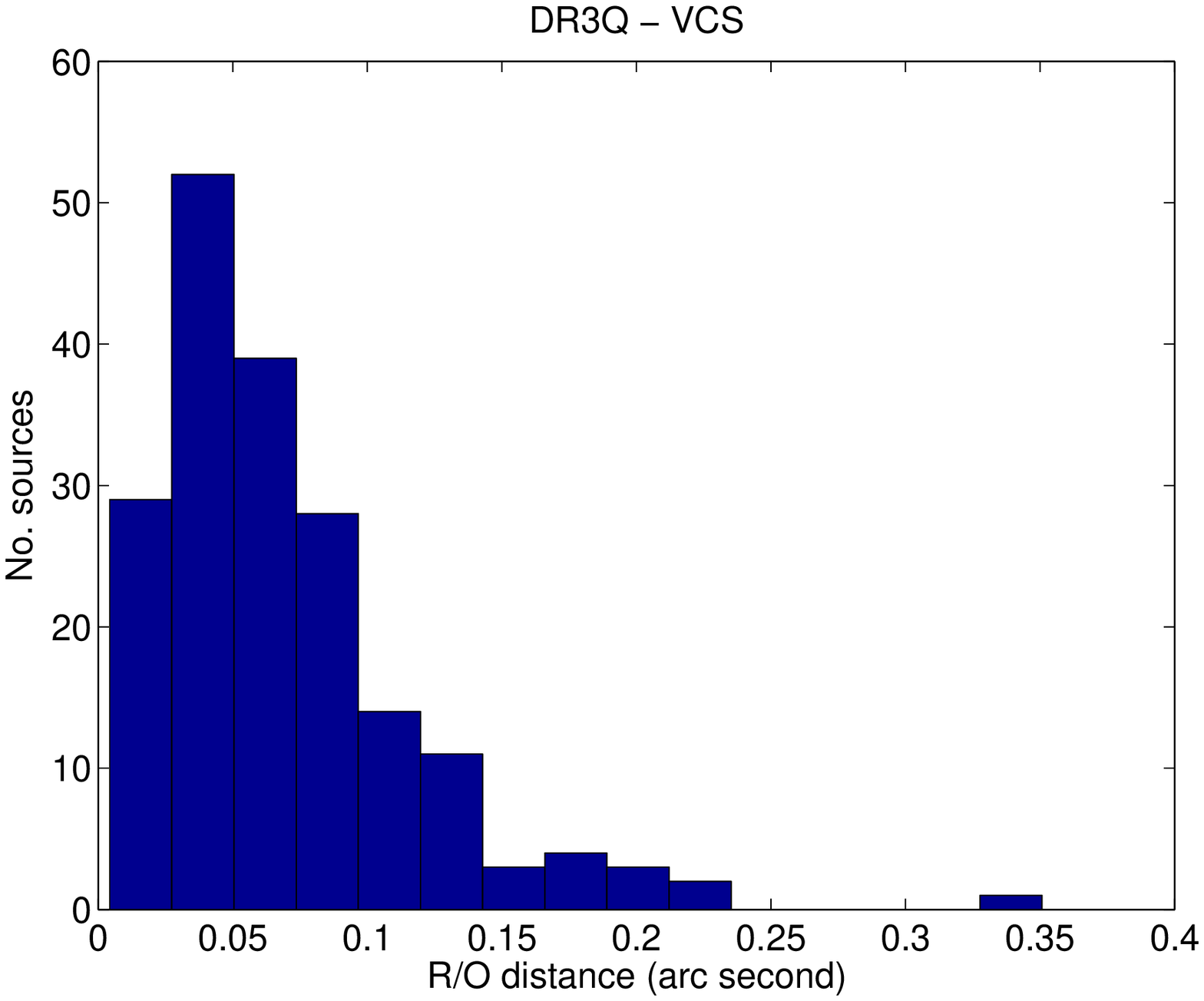}
\includegraphics[width=8cm]{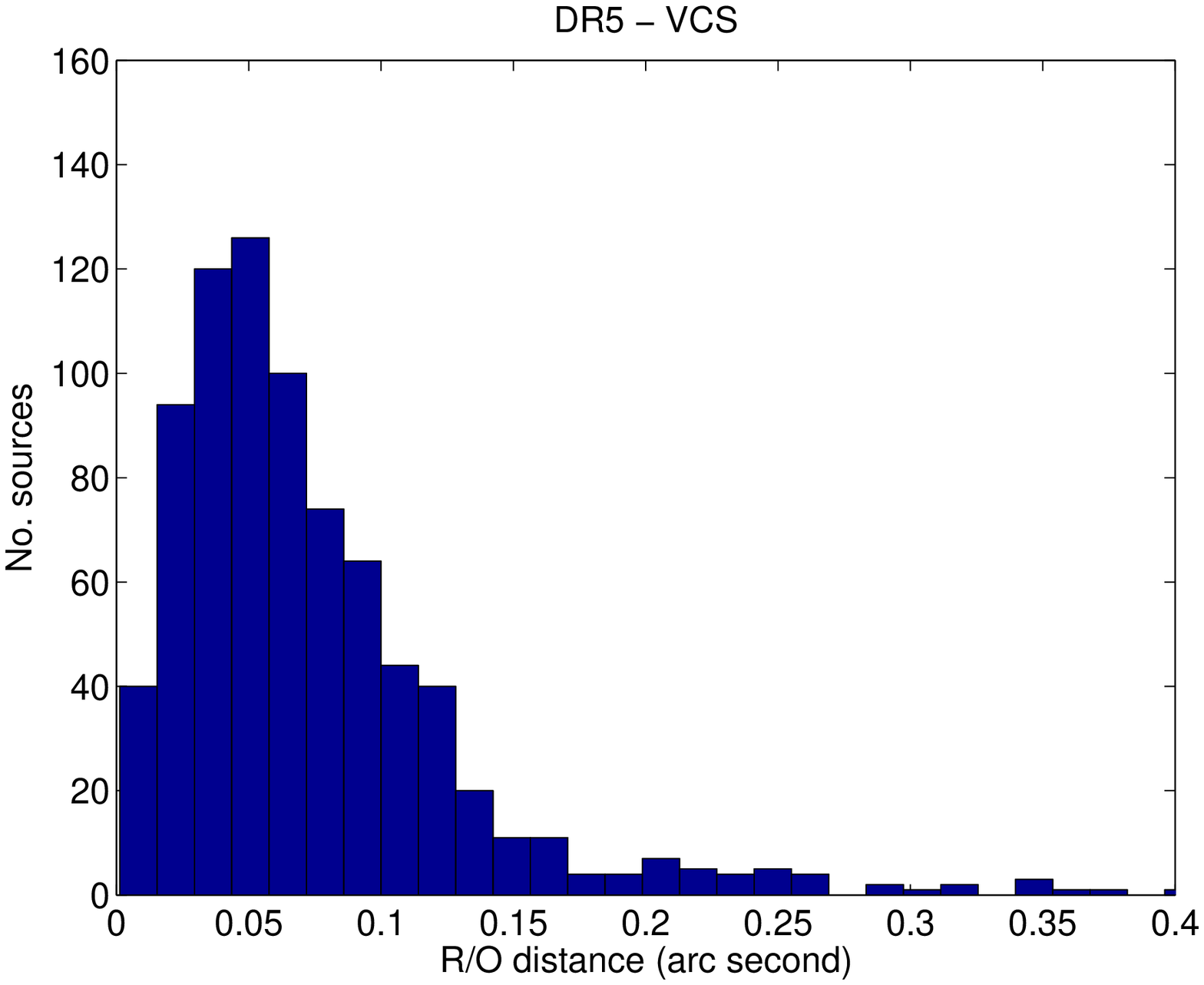}
\includegraphics[width=8cm]{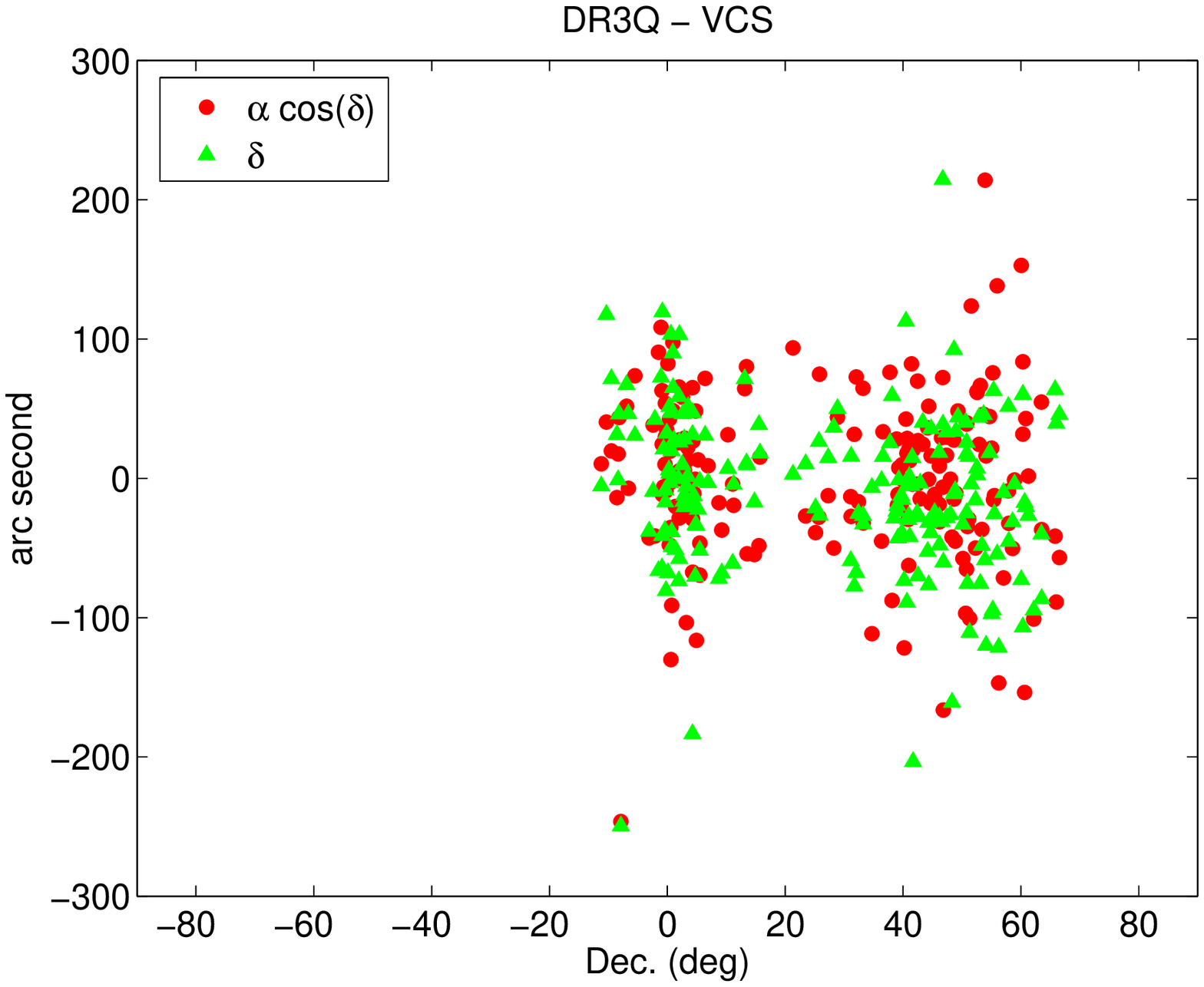}
\includegraphics[width=8cm]{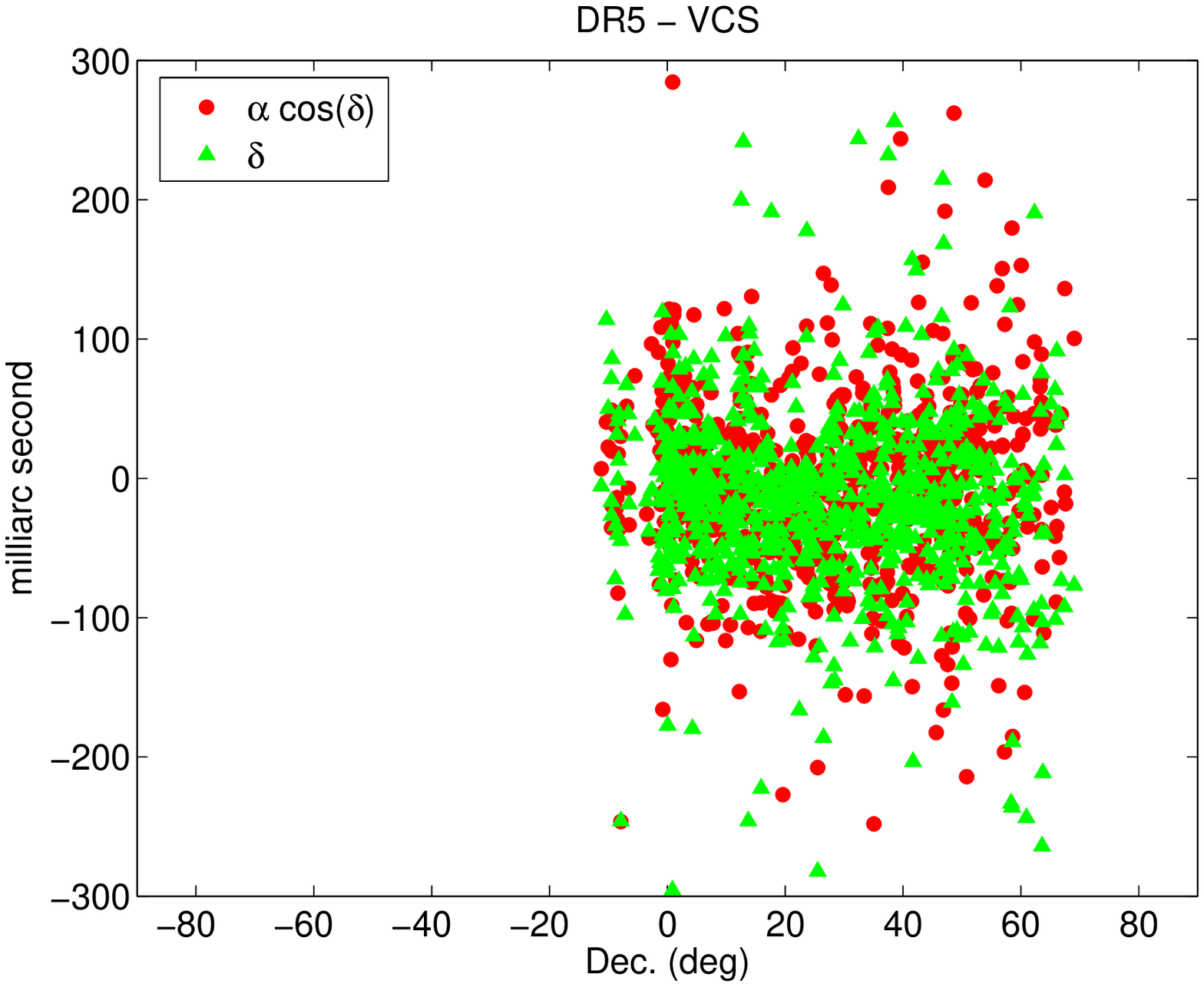}
\caption{Number of sources versus the R/O distance resulting from the cross-identification between the VCS and the DR3Q (left column) and the DR5 (right column) with a cut-off radius of 1~arc second and R/O distance in right ascension and declination versus the declination.}
\label{fig02}
\end{figure*}

\begin{table}
\caption[]{\small Statistics of the cross-identified sources. Unit:~mas.}
\label{stats}
\begin{center}
\begin{tabular}{lrrrr}
\hline
\hline
\noalign{\smallskip}
 & \multicolumn{2}{c}{DR3Q--VCS} & \multicolumn{2}{c}{DR5--VCS} \\
\noalign{\smallskip}  
\hline
\noalign{\smallskip}
cut-off radius   &    1~as & &   1~as &   \\
No. sources      &     186 & &    796 &   \\
\noalign{\smallskip}
\hline
\noalign{\smallskip}
& mean & wrms & mean & wrms \\
\noalign{\smallskip}
\hline
\noalign{\smallskip}
$\alpha$         &  $ -1.75$ &    86.11&   $   7.71$ & 107.16 \\
$\delta$         &  $  9.51$ &    57.99&   $  13.12$ &  80.95 \\
radius           &  $ 81.66$ &    64.84&   $  91.61$ &  99.37 \\
\noalign{\smallskip}
\hline
\end{tabular}
\end{center}
\end{table}

The cross-identification algorithm consists in checking whether an optical source of the DR3Q or the DR5 exists within a fixed cut-off radius around a given radio source of the VCS. The cut-off radius should be taken of the order of the astrometric precision of the less accurate catalogue. For instance, when comparing the VCS (typical accuracy: 1~mas) with the SDSS optical catalogue, for which the accuracy is close to 1~as, a wise cut-off radius will be at least the accuracy of the relevant optical catalogue. We adopt a cut-off radius of 1~as. We compute the difference of coordinates in the sense optical minus radio. The cross-identified sources are displayed on Fig.~\ref{fig01}. The repartition of the radio-optical (R/O) distances obtained after cross-identification between radio and optical catalogues is shown on Fig.~\ref{fig02}. The detailed statistics for each coordinate and for the radius are reported in Table~\ref{stats}.

The comparison DR3Q--VCS yields 186 sources. We find a mean R/O distance of 82~mas with a 65~mas wrms, corresponding to the broad peak on upper left the histogram on Fig.~\ref{fig02}. These values are consistent with the value of the SDSS astrometric accuracy estimated by Pier et al.~(2003) by comparison with other photometric catalogues. The mean R/O distance in declination admits a significantly larger mean value (10~mas) compared to the right ascension ($-2$~mas). This departure of the declination indicates a possible systematic shift of all the declinations and will be a key point in next section for the realization of a global reference frame.

\begin{figure}[htbp]
\centering
\includegraphics[width=8cm]{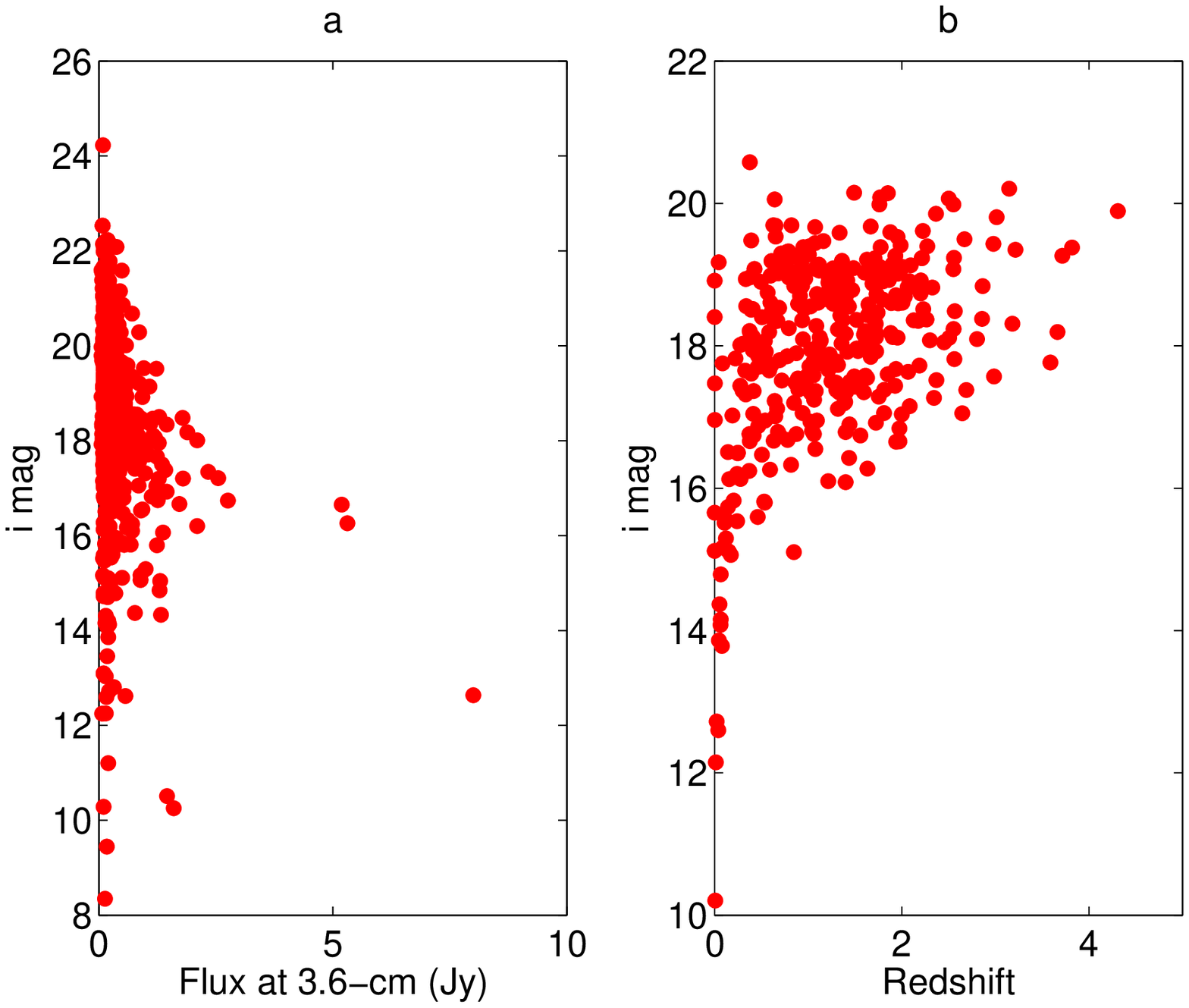}
\includegraphics[width=8cm]{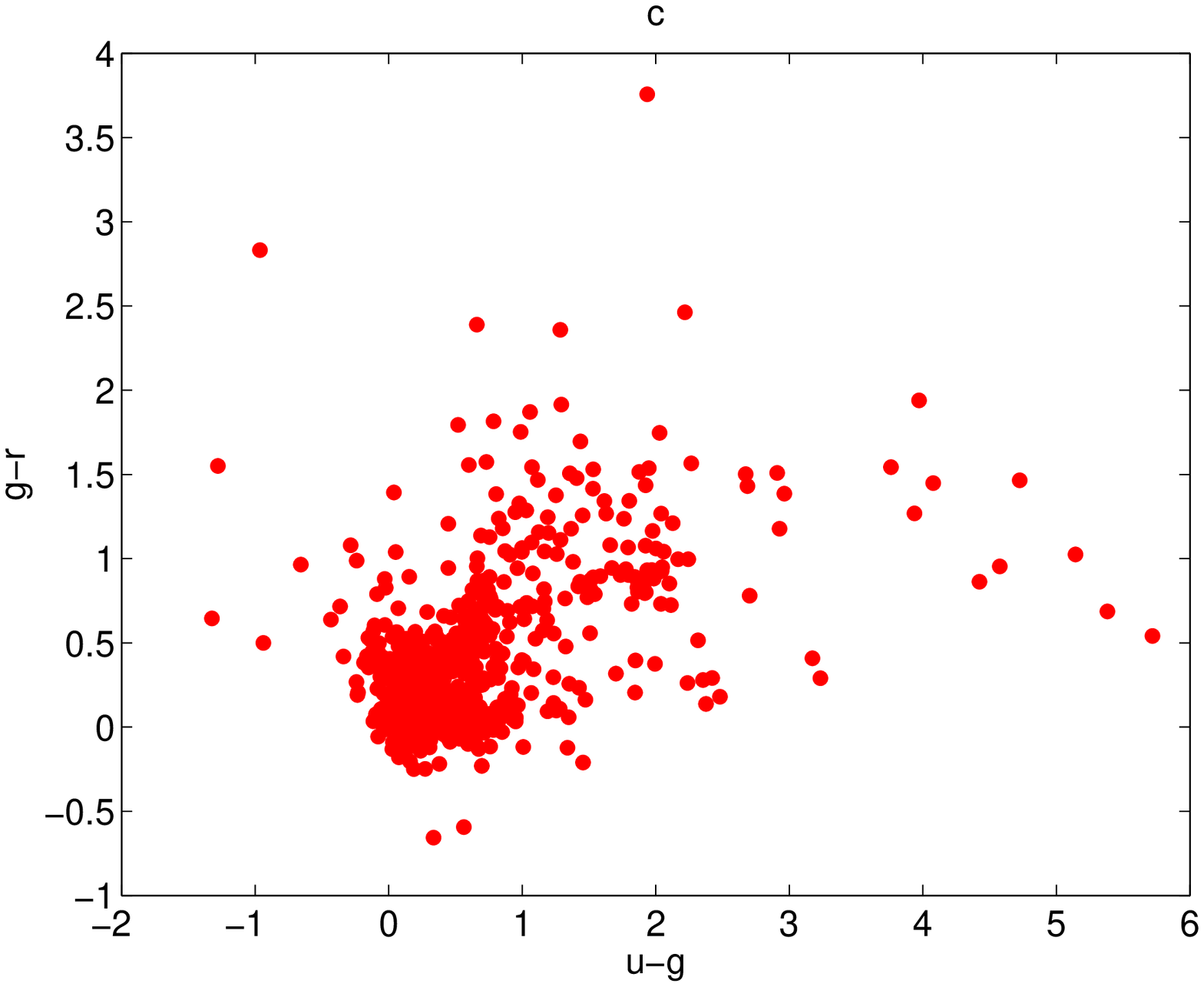}
\caption{Photometric properties of the cross-identified sources: (a) $i$ magnitude vs. the flux at 3.6-cm (X-band), (b) $i$ magnitude vs. the redshift, and (c) $g-r$ vs. $u-r$.}
\label{fig03}
\end{figure}

Note that 186 sources can appear as a strangely small value considering the several tens of thousands sources contained in the DR3Q. Indeed, the quasar selection algorithm of Schneider et al. (2005) can miss some quasars if they are too bright. This is for instance the case for the well-known quasar 3C~273 that is not selected although it is in the DR3Q region of the sky.

The comparison DR5--VCS returns 796 sources, which is about one fifth of the complete VCS data. This value is satisfying considering that the SDSS covers one quarter of the sky. We find a mean R/O distance of 92~mas with a 99~mas wrms (see also Fig.~\ref{fig02}). The mean R/O distance in declination admits also a mean value (13~mas) a bit larger than the right ascension one~(7~mas).

The R/O distances versus the declination are depicted on Fig.~\ref{fig02} (bottom parts). It appears that they are reparted uniformly through the covered span of declinations (roughly from $-10^{\circ}$ to $+60^{\circ}$), yielding that there is no visible effect of the declination on the R/O distance.

The cross-identification between the optical SDSS catalogues and the VCS radio catalogue allowed us to make interesting studies between the radio and optical properties of the 347 common sources for which the redshift is available. In Fig.~\ref{fig03}(a) we present the $i$~magnitude as a function of the flux at 3.6-cm wavelength. Notice the large global dispersion of the optical apparent brightness of the objects (recall that there is a 10,000 flux ratio between an object with $i=12$ and $i=22$) in comparison with the small dispersion of the radio flux. Fig.~\ref{fig03}(b) shows, as an expected result, that the $i$~magnitude of the objects tends to increase as a function of the redshift. A quasi linear threshold is suggested for the inferior values, which would mean that no quasar can reach a magnitude smaller than that given by this threshold at a given redshift. Fig.~\ref{fig03}(c) gives the color distribution $u-g$ as a function of $g-r$. We can notice that the SDSS/VCS sources are mainly extended along the $u-g$ scale.

\section{Global comparison of radio and optical coordinates}

We have two sets of coordinates ($\alpha$, $\delta$) for a number of sources indicated in Table~\ref{stats}, given in both radio and optical wavelengths. A question arises: how consistent are the axes and poles of the celestial reference frames realized separately through radio and optical sets of source coordinates? Since the two frames are realized with the same set of sources, we seek for the effect of the observing wavelength, for instance centimetric and optical wavelengths.

The coordinate difference is modeled by a global rotation around the three axes of the realized reference frame, represented by three angles $A_1$, $A_2$, $A_3$, and a bias $dz$ accounting for any systematic error in declination tilting the equator (see, e.g., Gontier et al.~2001):
\begin{eqnarray} \label{rot}
\Delta\alpha&=&A_1\tan\delta\cos\alpha+A_2\tan\delta\sin\alpha-A_3 \\
\Delta\delta&=&-A_1\sin\alpha+A_2\cos\alpha+dz
\end{eqnarray}

Note that a large value of the angle $A_3$ would indicate a rotation around the $Z$ axis and therefore a systematic shift of all the sources along the right ascension. Similarly, a significant value of $dz$ would mean a systematic shift of all the sources along the $Z$ axis, or equivalently a systematic offset in declination. Noticeable values for $A_1$ or $A_2$ would signify that both catalogues are rotated around the $X$ ($\alpha=0$ hour) and $Y$ ($\alpha=6$ hours) axes respectively, and have therefore different poles.

The estimation of the four unknown parameters is done over the cross-identified sources within the cut-off radius as defined in Table~\ref{stats} and using a least-squares fit. The difference of coordinates for each source is computed in the sense optical minus radio, so that the angles of rotation are determined for going from the radio reference frame towards its optical counterpart. Once the angles are determined, the rotations (1)--(2) is applied to the optical catalogue to correct all its source coordinates. Then, new rotation angles and bias are computed in the same way to check that there is no residual rotation. Results are displayed in Table~\ref{rotation}.

\begin{table}
\caption[]{\small Angles of rotation for the reference system realized through the cross-identified sources which statistics are reported in Table~\ref{stats}. Unit:~mas.}
\label{rotation}
\begin{center}
\begin{tabular}{lrcrrcr}
\hline
\hline
\noalign{\smallskip}
 & \multicolumn{3}{c}{DR3Q--VCS} & \multicolumn{3}{c}{DR5--VCS} \\
\noalign{\smallskip}
\hline
\noalign{\smallskip}
  $A_1$ & $  2.84$ & $\pm$ & $6.63$ & $  6.29$ & $\pm$ & $4.44$ \\
  $A_2$ & $-19.52$ & $\pm$ & $6.59$ & $-19.51$ & $\pm$ & $4.77$ \\
  $A_3$ & $ -0.28$ & $\pm$ & $6.28$ & $-10.22$ & $\pm$ & $3.98$ \\
  $dz$  & $ -0.49$ & $\pm$ & $6.36$ & $  0.44$ & $\pm$ & $4.52$ \\
\noalign{\smallskip}
\hline
\end{tabular}
\end{center}
\end{table}

Note that the formal error is comparable for all estimated parameters: $\sigma\sim5$~mas, the value being of the order of magnitude of the error ellipse of the VCS determined by Petrov et~al.~(2006).

Concerning the comparison DR3Q--VCS, it appears that the angles $A_1$ and $A_3$ and the bias quantity $dz$ are not statistically significant, considering their respective formal error. However, the angle $A_2$ around the $Y$ axis is significant ($3\sigma$). For the comparison DR5--VCS, $A_2$ is at almost $4\sigma$, and $A_3$ is roughly at $2\sigma$, whereas $A_1$ and $dz$ are not significant.

The angle $A_2$ is significantly larger than the formal error by at least a factor of three ($A_2=-20\pm7$~mas for both DR3Q--VCS and DR5--VCS). Theses values of $A_2$ point out a problem associated with the determination of the declination in one or both wavelengths. A~key point for the interpretation of this observation is the fact that most of the SDSS sources are located in about one quarter of the sky in the opposite direction to the $X$ axis (see Fig.~\ref{fig01}, the repartition of the SDSS cross-identified sources is mainly around $\alpha=12$~hours). A systematic error in the declination of these sources would appear as a pole offset in the direction of the $X$ axis, or equivalently, as a rotation around the $Y$ axis. Actually, a systematic shift of the declination generally shows up in the bias $dz$. In our case, it does not, but the correlation between $dz$ and $A_2$ is higher than 0.7 (whereas all other correlations are smaller by at least a factor of 10), indicating that $dz$ and $A_2$ are showing very intricated effects. If the repartition of the sources were uniform in right ascension, one would have expected a very small correlation between $dz$ and $A_2$ as well as a significant value for $dz$, whereas $A_2$ would have been expected to be non statistically significant. To confirm these conclusions, we generated a 'fake' optical catalogue by taking (i) the VCS radio coordinates of the 796 cross-identified sources with the DR5, and adding a random position with a systematic shift in declination, and (ii) the VCS radio coordinates of all sources (therefore covering the sky more uniformly) and adding a random position with the same systematic shift. In the case (ii), the $dz$ totally absorbs the offset in declination, while in the case (i), the offset appears in both $A_2$ and~$dz$.

Note that, after rotation of the optical catalogue, all of the angles are found null. The formal error stays at the same value than before the rotation.

\begin{figure}[htbp]
\centering
\includegraphics[width=8cm]{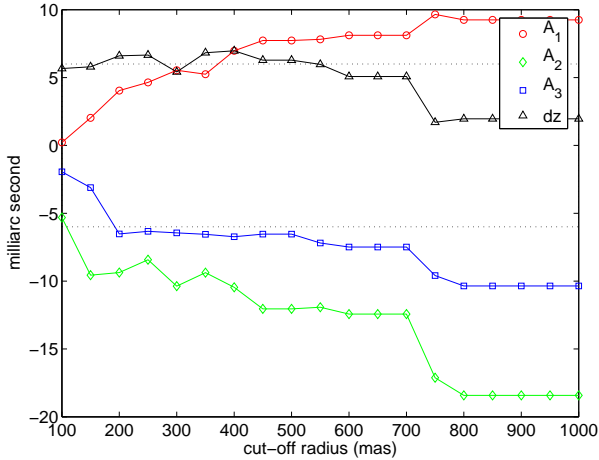}
\caption{Evolution of the rotation angles with the cut-off radius of cross-identification. The two horizontal dotted lines at $\pm6$~mas show the level of formal error associated to the determination of the angles.}
\label{fig04}
\end{figure}

The stability of the angles when the radius for the search of cross-identifications changes has been assessed. Fig.~\ref{fig04} displays the evolution of the four parameters when the radius runs from 0.1~as to 1~as. The main results associated to this plot are: (i) $A_1$ decreases with the cut-off radius, and becomes non significant when the cut-off radius passes below 0.4~as, (ii) $A_3$ looses its significance when the cut-off radius downs below 0.5~as, and (iii) $A_2$ and $dz$ are very correlated (as noticed in the previous paragraphs of this section), so that, when the cut-off radius decreases (in absolute value), $dz$ shows up. Note that, even though the cut-off radius is at 0.1~as, the number of cross-identified sources is still around 500, and therefore the statistics remain reliable.

The stability of the angles following the declination of the sources has been investigated too. Since the distribution of the DR5 cross-identified sources ranges from the equator to the mid-declinations, we computed the corresponding rotation angles between the DR5 and VCS two bands of declinations: below and above $20^{\circ}$~N. Results are reported in Table~\ref{bands}. They show that the angles are more sensitive to the sources with high declinations. This effect can be attributed to the distribution of the sources on the sky: sources with declinations higher than $20^{\circ}$~N are gathered between 8 and 18 hours in right ascension, whereas the rest of the sources covers more uniformly the right ascensions (see Fig.~\ref{fig01}).

\begin{table}
\caption[]{\small Angles of rotation between DR5 and VCS for two bands of declination. Unit:~mas.}
\label{bands}
\begin{center}
\begin{tabular}{lrcrrcr}
\hline
\hline
\noalign{\smallskip}
 & \multicolumn{3}{c}{$<$ 20$^{\circ}$ N} & \multicolumn{3}{c}{$>$ 20$^{\circ}$ N} \\
\noalign{\smallskip}
\hline
\noalign{\smallskip}
  $A_1$& $ 7.74$ & $\pm$ & $4.46$  & $  8.93$ & $\pm$ & $5.06$ \\
  $A_2$& $-8.42$ & $\pm$ & $3.36$  & $-27.53$ & $\pm$ & $5.84$ \\
  $A_3$& $-7.25$ & $\pm$ & $2.46$  & $-14.08$ & $\pm$ & $5.38$ \\
  $dz$ & $ 3.34$ & $\pm$ & $2.87$  & $ -4.40$ & $\pm$ & $5.87$ \\
\noalign{\smallskip}
\hline
\end{tabular}
\end{center}
\end{table}

For both the DR3Q--VCS and the DR5--VCS comparisons, the angle $A_2$ is negative. A negative $A_2$ shows that either the optical declinations are underestimated, or conversely, the radio declinations are overestimated. The question now arises about which one of the optical or the radio catalogue is responsible for this spurious offset. Gontier et al. (2001) computed time series of position of some ICRF radio sources using more than ten years of geodetic VLBI observations. It appeared that the variability of the source position is higher in declination than in right ascension for low declinations (i.e., sources close to the equator), although the correction for troposphere delay (including site-dependent troposphere gradients) was turned on. Deficiencies in the global mapping functions and the troposphere gradient modeling (MacMillan \& Ma 1997) could be at the origin of these effects. (Some promising issues on a different way to obtain local troposphere gradients have recently been carried out by B\"ohm \& Schuh (2006) but have not yet been applied to the computation of source positions.) However, the amplitude of the effect on the reference frame stability is expected to amount to a few tens of micro arc seconds and could not therefore account for the $A_2$ observed in this study. Pier et al. (2003) mentioned that additional systematic errors show up in the astrometric calibration of the SDSS due to anomalous refraction, random errors in the primary reference catalogues, or charge transfer efficiency effects in the astrometric CCD. The order of magnitude ($\sim$30~mas rms) of these effects could explain partly the observed systematic offset of the declinations.

\section{Discussion and conclusion}

This study shows that a comparison of existing radio and optical catalogues, respectively the fifth release of the VLBA Calibrator Survey (VCS) and the latest release of the SDSS (DR5), permits the cross-identification of 796 sources. Only 186 are cross-identified between the VCS and the SDSS DR3 quasar catalogue of Schneider et al.~(2005) due to the fact that the selection algorithm rejects some objects, although they are quasars (e.g., 3C~273). The accuracy of the SDSS is found around 100~mas, consistently with Pier et al. (2003). Celestial reference frames realized with the radio and optical coordinates of the cross-identified sources are consistent at the level of a few tens of mas. The main unconsistency appears as a~$-20\pm5$~mas global rotation around the $Y$ axis of the VCS sources with respect to the SDSS ones. This effect is statistically significant (almost $4\sigma$) compared to the VCS error ellipse (Petrov et al. 2005), and corresponds to an offset of the SDSS declinations with respect to the VCS ones. This effect shows up on the angle $A_2$ mainly because the SDSS covers a limited area centered in the direction of $-X$ ($\alpha=12$ hours). It is expected to disappear on $A_2$ and show up on $dz$ if the coverage in right ascension were uniform. Such an effect is likely due to a non-optimal astrometric calibration of the SDSS and should be addressed for future accurate tying of radio and optical astrometric data~sets.

In addition to these results our cross identifications allowed us to carry out interesting photometric studies linking optical characteristics of the common quasars (magnitude, color index, redshift) to their radio properties (flux). In particular we have remarked in our sample that the magnitude dispersion of the objects is relatively large compared to the radio flux dispersion.

As noticed by Souchay et al. (2006), the coverage for both physical and astrometric properties of all the ICRF objects in optical wavelengths is still unperfect. We encourage the SDSS and SDSS-like surveys to densify the observations of radio-observed objects in the future. For instance, in the SDSS data release 3 quasar catalogue we used in this study, a hundred of sources observed by the VLBA had not been observed by the SDSS although they are in the region of the sky covered by the latter survey. We emphasize also the necessity of covering all the right ascensions to avoid correlations in the determination of the transformation parameters between radio and optical frames.

An effort has also to be done on the side of VLBI and radio reference frames. The permanent geodetic VLBI network seems to be the best way to promote a regular monitoring of a large number of compact extragalactic objects. Some issues has to be mentioned. For instance, the variability of the network geometry from one to another VLBI observing sessions has significant effects on the determination of the Earth orientation (Lambert \& Gontier 2006). Problems linked to specific stations (e.g., size of the dish, locally strong troposphere gradients due to geographical features) or to the global network (e.g., length and orientation of the baselines) also influence the determination on the position of the observing sites (see the recent study of Feissel-Vernier et al. 2006). In a similar way, the influence of the network geometry and observing strategy on the realization of the celestial reference frame has to be carefully investigated.

Densification programs have also been undertaken to densify the ICRF in radio wavelengths in the southern hemisphere (Roopesh et al. 2005) together with observing programs at higher radio frequencies (Fey et al. 2005). Moreover, as noticed by Charlot (2004), the VCS is a corner-stone among the current densification programs. Nevertheless, the VLBA interferometer is a relatively small network compared to the intercontinental baselines used routinely in the geodetic VLBI networks (e.g., those used for the realization of the ICRF). Observing the VCS sources with a larger array could enforce the determination of their position. This could be realized by integrating progressively the VCS sources within the routine geodetic VLBI experiment schedule.

Finally, we would like to point out that such a study takes place naturally in the framework of the International Celestial Reference System Product Center (ICRS/PC), a joint collaboration between the United States Naval Observatory (Washington, DC) and the Paris Observatory (heads: R. Gaume and J. Souchay), which role is the monitoring of the ICRS, the maintenance of its current realization and the linking with other celestial reference frames, within the International Earth Rotation and Reference Systems Service~(IERS) and in coordination with the International VLBI Service for Geodesy and Astrometry~(IVS, Schl\"uter et al. 2002).

\begin{acknowledgements}
Funding for the Sloan Digital Sky Survey (SDSS) and SDSS-II has been provided by the Alfred P. Sloan Foundation, the Participating Institutions, the National Science Foundation, the United States Department of Energy, the National Aeronautics and Space Administration, the Japanese Monbukagakusho, the Max Planck Society, and the Higher Education Funding Council for England. The SDSS web site is http://www.sdss.org/. The SDSS is managed by the Astrophysical Research Consortium (ARC) for the Participating Institutions. The Participating Institutions are the American Museum of Natural History, Astrophysical Institute Potsdam, University of Basel, University of Cambridge, Case Western Reserve University, the University of Chicago, Drexel University, Fermilab, the Institute for Advanced Study, the Japan Participation Group, the John Hopkins University, the Joint Institute for Nuclear Astrophysics, the Kavli Institute for Particle Astrophysics and Cosmology, the Korean Scientist Group, the Chinese Academy of Sciences (LAMOST), Los Alamos National Laboratory, the Max-Planck-Institute for Astronomy (MPIA), the Max-Planck-Institute for Astrophysics (MPA), New Mexico State University, Ohio State University, University of Pittsburgh, University of Portsmouth, Princeton University, the United States Naval Observatory, and the University of Washington.
\end{acknowledgements}


\begin{thebibliography}{}

\bibitem{}Adelman-McCarthy, J.K., Ag\"ueros, M.A., Allam, S.S., et al. 2006, submitted to AJ Supp. Series

\bibitem{}B\"ohm, J., \& Schuh, H. 2006, In: D. Behrend and K.D. Baver (Eds.): International VLBI Service for Geodesy and Astrometry (IVS) 2006 General Meeting Proceedings, NASA/CP-2006-214140, 261

\bibitem{}Charlot, P. 2004, In: N.R. Vandenberg and K.D. Baver (Eds.): International VLBI Service for Geodesy and Astrometry (IVS) 2004 General Meeting Proceedings, NASA/CP-2004-212255, 12

\bibitem{}Charlot, P., Fey, A., Ojha, R., \& Boboltz, D. 2006, In: International VLBI Service for Geodesy and Astrometry (IVS) 2006 General Meeting Proceedings, D. Behrend and K.D. Baver (Eds.), NASA/CP-2006-214140, 321

\bibitem{}Danner, R., Unwin, S., \& Allen, R.J. 1999, SIM: Space Interferometry Mission: Taking the Measure of the Universe (Washington, DC: NASA)

\bibitem{}Dehant, V., Feissel-Vernier, M., de Viron, O., et al. 2003, J. Geophys. Res., 108(B5), 10.1029/2002JB001763

\bibitem{}Feissel, M., Gontier, A.-M., \& Eubanks, T.M. 2000, A\&A, 359, 1201

\bibitem{}Feissel-Vernier, M. 2003, A\&A, 403, 105

\bibitem{}Feissel-Vernier, M., Ray, J., Altamimi, Z., et al. 2004, In: N.R. Vandenberg and K.D. Baver (Eds.): International VLBI Service for Geodesy and Astrometry (IVS) 2004 General Meeting Proceedings, NASA/CP-2004-212255, 22

\bibitem{}Feissel-Vernier, M., Ma, C., Gontier, A.-M., \& Barache, C. 2005, A\&A, 438, 1141

\bibitem{}Feissel-Vernier, M., Ma, C., Gontier, A.-M., \& Barache, C. 2006, A\&A, 452, 1107

\bibitem{}Feissel-Vernier, M., de Viron, O., \& Le Bail, K. 2006, in preparation for Earth Planets Space

\bibitem{}Fey, A.L., Eubanks, T.M., \& Kingham, K.A. 1997, In: American Astronomical Society, 191st AAS Meeting, \#22.13; Bulletin of the American Astronomical Society, 29, 1249

\bibitem{}Fey, A.L., Boboltz, D.A., \& Charlot, P. 2005, In: J. Romney and M. Reid (Eds.): Future Directions in High Resolution Astronomy: The 10th Anniversary of the VLBA, ASP Conference Proceedings, 340, 514

\bibitem{}Fukugita, M., Ichikawa, T., Gunn, J.E., et al. 1996, AJ, 111(4), 1748

\bibitem{}Gontier, A.-M., Le Bail, K., Feissel, M., \& Eubanks, T.M. 2001, A\&A, 375, 661

\bibitem{}Herring, T.A., Mathews, P.M., \& Buffet, B.A. 2002, J. Geophys. Res., 107(B4), DOI 10.1029/2001JB000165 

\bibitem{}Kovalev, Y.Y., Petrov, L., Fomalont, E., \& Gordon, D. 2006, in preparation for AJ

\bibitem{}Lambert, S.B., \& Gontier, A.-M. 2006, In: D. Behrend and K.D. Baver (Eds.): International VLBI Service for Geodesy and Astrometry (IVS) 2006 General Meeting Proceedings, NASA/CP-2006-214140, 264 

\bibitem{}Lambert, S.B. 2006, A\&A, 457, 717

\bibitem{}Ma, C., Arias, E.F., Eubanks, T.M., et al. 1998, AJ, 116, 516

\bibitem{}MacMillan, D.S., \& Ma. C. 1997, Geophys. Res. Lett., 24, 453

\bibitem{}Mathews, P.M., Herring, T.A., \& Buffett, B.A. 2002, J. Geophys. Res., 107(B4), 2068, doi:10.1029/2001JB000390

\bibitem{}Mignard, F. 2003, In: R. Gaume et al. (Eds.): IAU XXV, Joint Discussion 16: The International Celestial Reference System, Maintenance and Fugure Realizations, 133

\bibitem{}Perryman, M.A.C., de Boer, K.S., Gilmore, G., et al. 2001, A\&A, 369, 339

\bibitem{}Petrov, L., Kovalev, Y.Y., Fomalont, E.B., \& Gordon, D. 2006, AJ, 131, 1872

\bibitem{}Pier, J.R., Munn, J.A., Hindsley, R.B., et al. 2003, AJ, 125, 1559

\bibitem{}Roopesh, O., Fey, A.L., Charlot, P., et al. 2005, AJ, 130(6), 2529

\bibitem{}Schl\"uter, W., Himwich, E., Nothnagel, A. et al. 2002, Adv. Sp. Res., 30(2), 145

\bibitem{}Schneider, D.P., Hall, P.B., Richards, G.T., et al. 2005, AJ, 130, 367

\bibitem{}SDSS DR5 2006, The SDSS Data Release 5, Description and data available at: http://www.sdss.org/dr5

\bibitem{}Souchay, J., Gontier, A.-M., \& Barache, C. 2006, A\&A, 453, 743


\bibitem{}Vondr\'ak, J., Ron, C., \& Weber, R. 2005, A\&A, 444, 297

\end{thebibliography}
\end{document}